\documentclass[12pt]{article}
\usepackage{epsfig}

\def\beq{\begin{equation}}
\def\eeq#1{\label{#1}\end{equation}}
\def\eeqn{\end{equation}}

\def\beqa{\begin{eqnarray}}
\def\eeqa#1{\label{#1}\end{eqnarray}}
\def\eeqan{\end{eqnarray}}

\let\bar=\overbar

\def\Dslash{\not{\hbox{\kern-4pt $D$}}}
\def\dslash{\not{\hbox{\kern-2pt $\del$}}}

\def\msb{{\bar{\ssstyle M \kern -1pt S}}}

\newcommand{\gtrsim}{\raisebox{-.25em}{$\stackrel{\mbox{\normalsize $>$}}{\sim}$}}
\newcommand{\lesssim}{\raisebox{-.25em}{$\stackrel{\mbox{\normalsize $<$}}{\sim}$}}

\newcommand{\bnP}{\bar {\cal P}}

\newcommand{\cP}{{\cal P}}
\newcommand{\bn}{{\bar n}}

\def\Dslash{D\!\!\!\!\slash}

\def\bnslash{\bar n\!\!\!\slash}

\def\Title#1{\begin{center} {\Large {\bf #1} } \end{center}}

\begin{document}

\vbox{\hbox{INT-PUB-02-42}} 

\Title{The Soft-Collinear Effective Field Theory\footnote{Invited plenary talk
at {\it Flavor Physics and CP Violation}, Philadelphia, PA, 2002.}}



\begin{raggedright}  

{\it Iain W. Stewart\index{Stewart, I.}\\
INT, Department of Physics\\
University of Washington\\
Box 351550\\
Seattle, WA 98195}
\bigskip\bigskip
\end{raggedright}
\vspace{-1cm}

\section{Introduction}

In general, to understand hadronic uncertainties we need a
separation of short $p\sim Q$ and long $p\sim \Lambda_{\rm QCD}$ distance
fluctuations. For processes in QCD with momentum transfers $Q\gg \Lambda_{\rm
QCD}$, the short distance part is calculable in terms of Wilson coefficients or
hard scattering functions. The long distance contributions can be arranged into
universal non-perturbative matrix elements which can be extracted from data or
calculated on the lattice. This process of separating short and long distance
fluctuations is sometimes referred to as factorization.

In $B$ physics a proper understanding of hadronic uncertainties is crucial,
partially since the size of typical expansion parameters $\Lambda_{\rm
QCD}/m_b$, $\Lambda_{\rm QCD}/m_c$, or $\sqrt{\Lambda_{\rm QCD}/m_b}$ leaves
room for power corrections to play a non-negligible role.  For many processes
Heavy Quark Effective Theory (HQET) provides the conceptual framework for
quantifying the factorization between the scales $m_{b,c}$ and $\Lambda_{\rm
QCD}$. A well known example is the extraction of the CKM matrix element
$|V_{cb}|$ from knowledge about the form factors in $B\to D^{*}\ell \nu$
decays. At zero recoil the leading result is fixed by Heavy Quark
Symmetry~\cite{IW}, and the first power corrections vanish~\cite{Luke}. More
recently important progress has been made at reducing the dominant model
dependence by computing the matrix elements of $1/m_Q^2$ operators on the
Lattice~\cite{Andreas}.  For inclusive $B\to X_c\ell\nu$ and $B\to X_u\ell\nu$
decays the non-perturbative HQET matrix elements $\overline\Lambda$ and
$\lambda_{1}$ are actively being extracted from experimental
data~\cite{Ecklund}.

Since the $B$ is so heavy, many of its decays produce energetic light hadrons.
For these decays the energy of the hadron $E_H$ in the $B$ rest frame is an
additional perturbative scale, and HQET alone does not separate the perturbative
and non-perturbative information. Examples of such processes include the decays
$B\to D\pi$, $B\to \pi\pi$, $B\to K\pi$, the large recoil region in
$B\to\pi\ell\nu$, $B\to \rho\ell \nu$, $B\to K^*\gamma$, and $B\to
K\ell^+\ell^-$, and the endpoint spectra of the inclusive decays $B\to X_u
\ell\nu$ and $B\to X_s\gamma$. The nature of factorization in these decays
shares features in common with many exclusive and inclusive hard scattering
processes. Examples are $\gamma^*\gamma\to \pi^0$ at large $q^2$ and the $x\sim
1$ endpoint region of deep inelastic scattering.

In this talk I discuss an effective field theory that has been developed for
processes with energetic hadrons, which is referred to as the Soft-Collinear
Effective Theory (SCET)~\cite{bfl,bfps,cbis,bpssoft}. SCET can be used for both
hard scattering processes~\cite{bfprs} and B decays. This theory makes
symmetries relevant in the large energy limit explicit at the level of the
Lagrangian and operators (such as the reduction of spin structures, helicity
constraints, and collinear gauge invariance). Furthermore, SCET has a
transparent power counting in $\lambda=\Lambda_{\rm QCD}/Q$ (or
$\lambda=\sqrt{\Lambda_{\rm QCD}/Q}$) so that power corrections can be
investigated in a systematic way~\cite{chay1,bpspc,beneke}. This includes
processes not amenable to an operator product expansion such as exclusive
decays. The renormalization group improvement of operators in the effective
theory sums single infrared logs, as well as double Sudakov logarithms when they
appear~\cite{bfl,bfps,bauer}.  Finally, SCET allows proofs of factorization
theorems to be simplified and carried out in a gauge invariant way.

\section{Formalism}

\begin{table}[t!]
\begin{center}
\begin{tabular}{cl|c|clc}
\hline\hline
  Type & Momenta $(+,-,\perp)$\hspace{0.4cm} 
   & \hspace{0.2cm}Field Scaling  \hspace{0.cm} 
   & \hspace{0.2cm}Operators\hspace{0.2cm} \\ 
   \hline
  collinear & $p^\mu\sim (\lambda^2,1,\lambda)$ \hspace{0.2cm}
   & \hspace{0.2cm} $\xi_{n,p}\sim \lambda$ & $\bnP$, $W_n$ $\sim\lambda^0$ \\
  && ($A_{n,p}^+$, $A_{n,p}^-$, $A_{n,p}^\perp$) $\sim$ 
  ($\lambda^2$,$1$,$\lambda$) & \hspace{0.4cm} $\cP_\perp^\mu\sim \lambda$ \\
   \hline
  soft &  $p^\mu\sim (\lambda,\lambda,\lambda)$ \hspace{0.25cm}
   & \hspace{0.8cm} $q_{s,p}\sim \lambda^{3/2}$ & \hspace{0.5cm} 
   $S_n \sim \lambda^0$ \\
  & & \hspace{0.2cm} $A_{s,p}^\mu\sim \lambda$ & \hspace{0.3cm} 
    $\cP^\mu\sim \lambda$ \\ \hline 
  usoft &  $k^\mu\sim (\lambda^2,\lambda^2,\lambda^2)$
   & \hspace{0.5cm} $q_{us}\sim \lambda^3$ & \hspace{0.4cm} $Y_n\sim\lambda^0$ \\
  & & \hspace{0.4cm} $A_{us}^\mu\sim \lambda^2$  \\
\hline\hline
\end{tabular}
\end{center}
\caption{\setlength\baselineskip{12pt} Power counting for SCET momenta and
fields as well as momentum label operators ($\bnP,\cP_\perp^\mu$, $\cP^\mu$) and
collinear and soft Wilson lines induced by integrating out offshell
fluctuations ($W$, $S_n$) and the usoft Wilson line $Y_n$ induced by a collinear
field redefinition as described in the text.  \label{table_pc}
\setlength\baselineskip{18pt}}
\end{table}
The factorization of scales in the effective theory is carried out by describing
long distance fluctuations with $p^2\,\lesssim\, Q^2\lambda^2$ using effective
theory fields, and those with $p^2\gg Q^2\lambda^2$ by computable short distance
Wilson coefficients. Typical processes require collinear fields and in addition
either soft or usoft fields. Examples are $B\to X_s\gamma$ which needs collinear
and usoft fields, and $B\to D\pi$ which needs collinear and soft fields.  This
field content is summarized in Table~\ref{table_pc} together with the scaling of
the momenta and fields with the expansion parameter $\lambda$~\cite{bfl,bfps}.
The momenta scales $Q$, $Q\lambda$, and $Q\lambda^2$ are separated by making
phase redefinitions to pull out the larger momenta, $\phi_n(x) =\sum_p
e^{-ip\cdot x}
\phi_{n,p}(x)$. Derivatives on the new fields then always pick out the small
scale, $\partial^\mu\phi_{n,p}(x)\sim (Q\lambda^2) \phi_{n,p}(x)$, while the
large momenta are picked out by introducing label operators, for example
$\bnP\, \xi_{n,p}=(\bn\!\cdot\!  p)\,\xi_{n,p}$. Since $\bnP\sim \lambda^0$ in
the power counting the hard coefficients $C(\bnP)$ are arbitrary functions of
this operator~\cite{cbis}, which can be determined by matching. More generally
we have functions $C(\omega_i) \prod_i \delta(\omega_i-\bnP)$ where the delta
functions are inserted inside collinear operators in the most general locations
allowed by gauge and reparameterization invariance.

Furthermore, there are gluon fields which are order $\lambda^0$ in the power
counting, namely $\bn\cdot\! A_{n,q}\sim 1$. Integrating out offshell
fluctuations builds up Wilson lines in these fields, such as in the
example~\cite{bfps} of matching the full theory heavy-to-light current $\bar u
\Gamma b$ onto the SCET current, $J_0=C(\bnP)\, \bar\xi_{n,p}W\Gamma h_v$.  In
fact the $\bn\cdot\! A_{n,q}$ field can be traded for the Wilson line
\begin{eqnarray}
  W &=& \Big[ \sum_{\rm perms} \exp\Big( -\frac{g}{\bnP}\: \bn\cdot A_{n,q}(x) \
  \Big) \Big] \,,
\end{eqnarray}
since the covariant derivative $i\bn\cdot D_c=\bnP+g\bn\cdot A_{n,q} = W\,
\bnP\, W^\dagger$. 
\begin{figure}[t!]
\begin{center}
 \epsfig{file=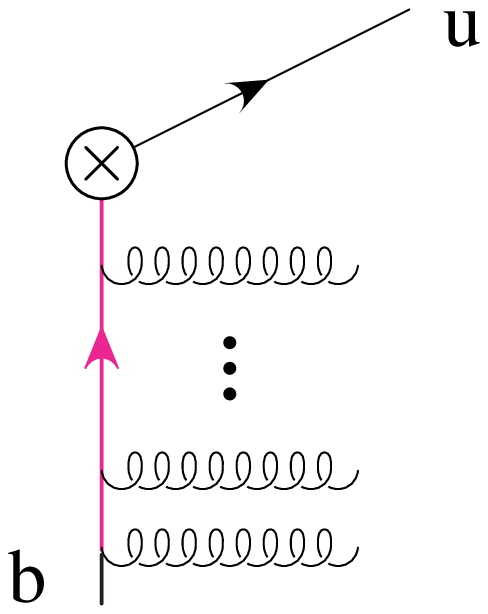,height=1.2in} 
 \hspace{0.7cm} \raisebox{1.4cm}{\Huge $\Longrightarrow$} \hspace{0.7cm}
 \epsfig{file=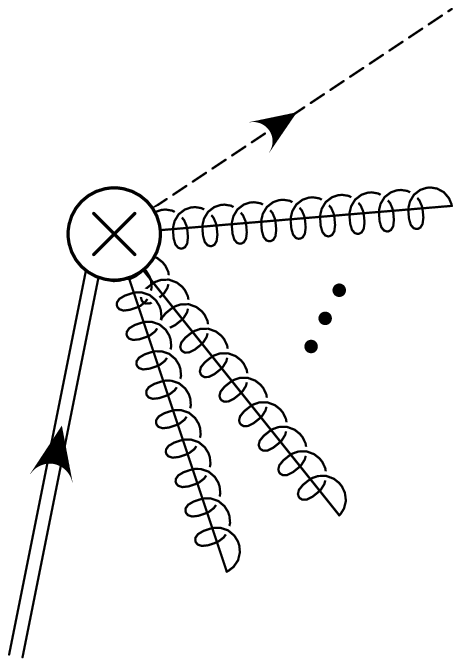,height=1.2in}
\caption{Tree level matching for the leading order heavy-to-light current}
\label{fig:match}
\end{center}
\end{figure}
Soft Wilson lines $S_n[n\cdot A_s]$ are also built up by integrating out
offshell fluctuations~\cite{bpssoft}.  Beyond tree level the structure of
operators containing factors of $W$ or $S_n$ is protected by collinear and soft
gauge transformations~\cite{cbis,bpssoft}.

In general we have three types of gluon fields, collinear, soft, and
usoft. These are the fields associated with gauge transformations $U(x)$ which
have support over collinear, soft, and usoft momenta
respectively~\cite{bpssoft}. The usoft fields $A_{us}^\mu$ are dynamical quantum
fields which due to their slow variation appear as background fields to the soft
and collinear quarks and gluons. For a gauge transformation with support over
collinear momenta it is convenient to factor out the large momentum
components, $U(x) = \sum_R e^{-i R\cdot x}\, {\cal U}_R(x)$. The collinear gluon
field then transforms such that
\begin{eqnarray} \label{A1}
 i{\cal D}^\mu + A_{n,p}^\mu \to {\cal U}_Q A_{n,R}^\mu {\cal U}^\dagger_{Q+R-p}
 + \frac{1}{g}\, {\cal U}_Q i{\cal D}^\mu {\cal U}^\dagger_{Q-p}  \,,
\end{eqnarray}
where $i{\cal D}^\mu = n^\mu \bnP/2 + \cP^\mu_\perp + \bn^\mu in\cdot D/2$.  On
the other hand an usoft gauge transformation has $i\partial^\mu U(x)\sim
(Q\lambda^2)U(x)$. For a fixed $R$ the transformation ${\cal U}_R(x)$ itself has
an usoft component (having no large phase), so there is an overlap between what
is meant by collinear and usoft gauge transformations. Effectively therefore one
also has
\begin{eqnarray} \label{Aus2}
 A_{us}^\mu \to {\cal U}_R A_{us}^\mu {\cal U}^\dagger_R
  + {\cal U}_R\, \frac{i}{g} \partial_\mu\, {\cal U}^\dagger_R \,.
\end{eqnarray}
This transformation is usoft -- it does not induce large momentum in the usoft
gauge field because the large momenta labels in ${\cal U}_R$ and ${\cal
U}_R^\dagger$ cancel.  Gauge invariance constrains the form of the Lagrangian at
leading and subleading orders. Beyond leading order the fact that the
transformations can not be completely separated is crucial to correctly
constrain operators as pointed out in Ref.~\cite{chay2}. For instance, the
collinear gauge invariance of the subleading term $\bar\xi_{n,p}
i\Dslash_{us}^\perp W (1/\bnP) W^\dagger i\Dslash_c^\perp \bnslash \xi_{n,p'}$
relies on the transformation in Eq.~(\ref{Aus2}). For the explicit form of the
leading order gluon and quark actions ${\cal L}_c^{(0)}$ we refer the reader to
Ref.~\cite{bpssoft}, for higher order terms in the collinear quark action to
Refs.~\cite{chay1,mmps}, and for the mixed collinear-usoft quark action to
Ref.~\cite{beneke}.
 
Besides the constraints from gauge invariance on collinear operators there are
addition constraints from the way in which Lorentz invariance is
realized in the effective theory. The collinear fields are defined by
introducing two auxillary light-like vectors, $n$ and $\bn$, such that $n\cdot
\bn=2$. Naively the presence of these vectors breaks Lorentz invariance.
However, in practice Lorentz invariance is restored order by order in the power
counting by a reparameterization invariance (RPI)~\cite{LM}. For the collinear
theory the study of RPI was initiated in Ref.~\cite{chay1} and generalized to
the three most general classes of allowed transformations in
Ref.~\cite{mmps}. For Lagrangians and operators with collinear fields, RPI gives
non-trivial constraints between the Wilson coefficients which appear different
orders in the power expansion. In general there is no way of deducing these
constraints using only the full theory.

Finally, it is worth discussing why the proof of factorization theorems is
simplified by using the effective theory. The factorization between hard and
collinear fluctuations or soft and collinear fluctuations is simplified by the
fact that it takes place at the level of matching onto the effective theory. The
resulting structures are constrained by the symmetries of the low energy theory
as already discussed. The factorization between collinear and usoft interactions
is simplified by the fact that many cancellations occur in a universal way at
the level of the effective Lagrangian. For instance, at lowest order the actions
for usoft and collinear particles can be factorized by a simple field
redefinition on the collinear fields, $\xi_{n,p}=Y_n
\xi_{n,p}^{(0)}$ and $A_{n,p} = Y_n A_{n,p}^{(0)} Y_n^\dagger$, where $Y_n =
P\exp [ig \int_{-\infty}^x\!\! ds\, n\!\cdot\! A_{us}(s n)]$. This
transformation moves all leading order usoft interactions from the collinear
Lagrangian ${\cal L}^{(0)}_c$ into the external operators and currents, after
which cancellations due to the unitarity of the usoft Wilson line, $Y_n^\dagger
Y_n=1$, are more readily seen~\cite{bpssoft}.

\section{Results}

I briefly discuss two B-physics applications of SCET, the exclusive decay $B\to
D\pi$ and the inclusive process $B\to X_s\gamma$.

For the decay $B\to D\pi$ the energy of the outgoing pion in the rest frame of
the $B$ is $E_\pi=2.31\,{\rm GeV}$. Since this energy is large it is useful to
consider this decay as being in the situation where $Q\gg \Lambda_{\rm QCD}$
with $Q=m_b$, $m_c$, or $E_\pi$. In this limit, SCET has been used to prove the
following factorization theorem~\cite{bps}
\begin{eqnarray} \label{fact1}
\langle D^{(*)} \pi | H_w  | B\rangle = N { F^{B\to D^{(*)}}(0)}\ 
 \int_0^1 d\xi\ { T(\xi,Q,\mu)}\ 
 {\phi_\pi(\xi,\mu)} + \ldots \,,
\end{eqnarray}
where the ellipses denote terms that vanish faster than the leading term as
$Q\to\infty$.  
\begin{figure}[t!]
\begin{center}
 \epsfig{file=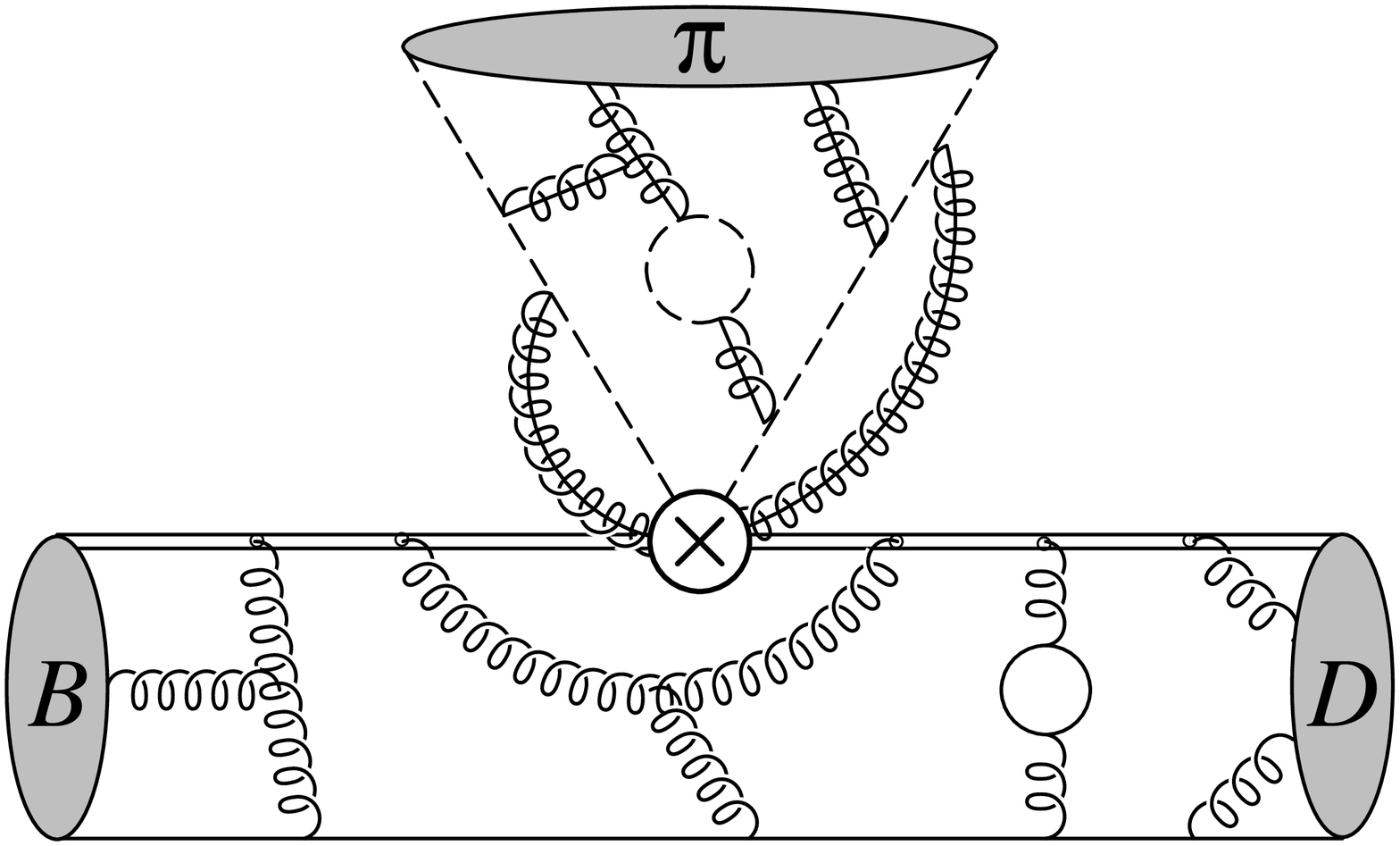,height=1.5in} \hspace{1.3cm}
  \epsfig{file=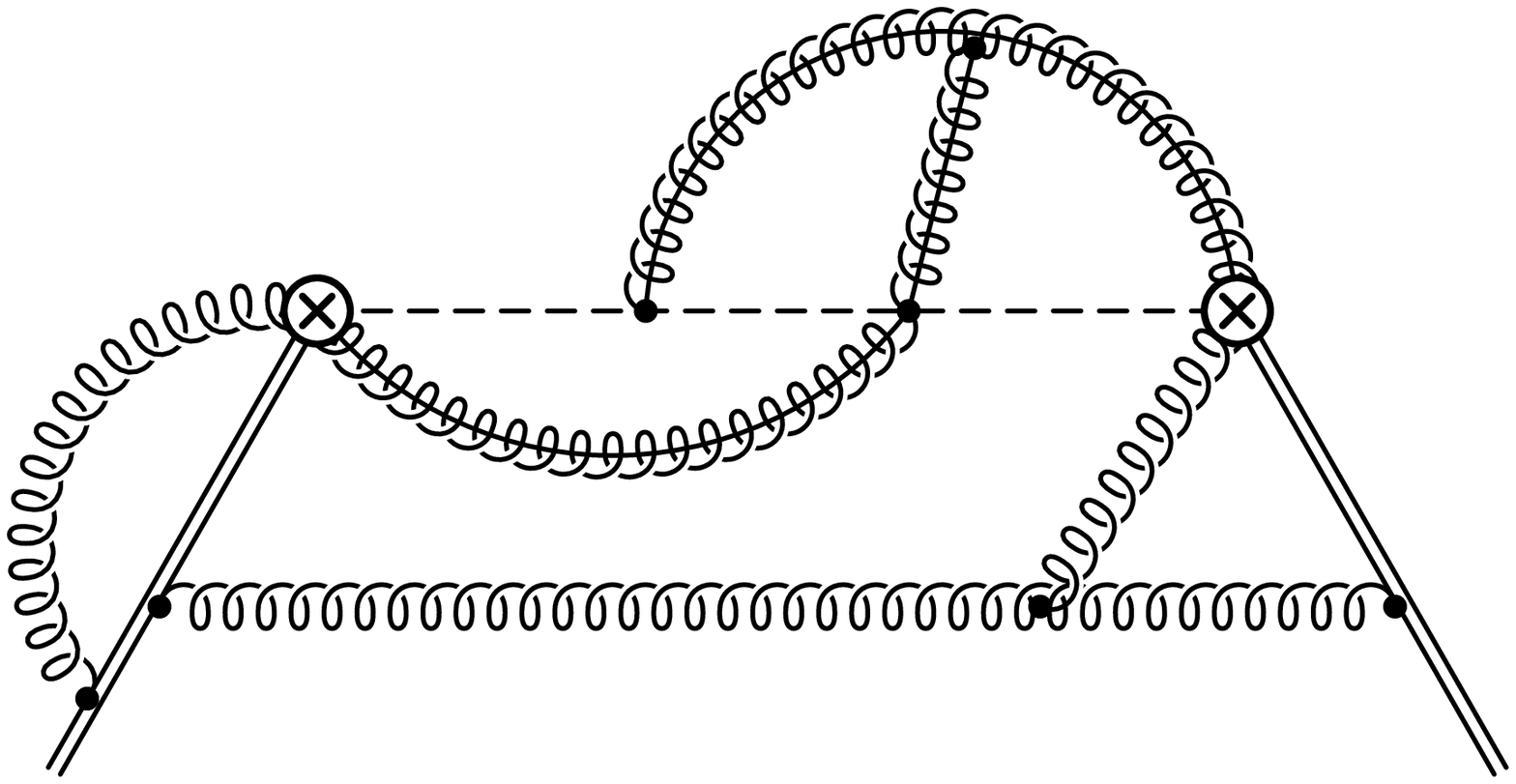,width=2.4in}
\caption{How the factorization of interactions occurs in 
$B\to D\pi$ and $B\to X_s\gamma$. Gluons with a line through them are collinear,
while those without are soft or usoft.}
\label{fig:bdpi}
\end{center}
\end{figure}
The relevant terms in the electroweak Hamiltonian are $H_W = C_1 O_1 + C_8 O_8$,
where $O_{1}=\bar d_L \gamma^\mu u_L\bar c_L \gamma_\mu b_L$ and $O_{8}= \bar
d_L T^A\gamma^\mu u_L \bar c_L T^A\gamma_\mu b_L$.  Eq.~(\ref{fact1}) was
proposed in Ref.~\cite{PW}, proven to two-loops in Ref.~\cite{bbns}, and proven
to all orders in $\alpha_s$ in Ref.~\cite{bps}.  The idea behind the proof is
shown in Fig.~\ref{fig:bdpi}. After integrating out offshell fluctuations the
leading order interactions involve soft gluons exchanged between quarks in the
$B$ and $D$ which build up the $B\to D$ form factor $F^{B\to D}$, and collinear
gluons exchanged between quarks in the pion building up the light-cone pion
wavefunction $\phi_\pi(x)$.  It should be remarked that the effective theory
analysis is carried out at an operator level so it does not rely on perturbation
theory.

For the proccess $B\to X_s\gamma$ in the region of photon energies
$E_\gamma\gtrsim m_B/2-\Lambda_{\rm QCD}\simeq 2.2\,{\rm GeV}$ the particles in
$X_s$ are collimated in a collinear jet with offshellness $p_X^2\simeq
m_B\Lambda_{\rm QCD}$. In this process the $B$ meson is dominated by usoft
dynamics and the leading power prediction from factorization is~\cite{KS}
\begin{eqnarray} \label{fact2}
  \frac{1}{\Gamma_0} \frac{d\Gamma}{dE_\gamma}
 = { H(m_b, \mu)}\int_{2E_\gamma-m_b}^{\bar \Lambda} \!\!\!\! dk^+\: 
   {S(k^+,\mu)}\:{ J(k^+\!+\! m_b \!-\! 2E_\gamma,\mu)} \,.
\end{eqnarray}
The SCET has been used to give a simple direct proof of this
result~\cite{bpssoft}. After the collinear field redefinitions, the usoft
interactions rearrange themselves to leave only diagrams such as the one shown
in Fig.~\ref{fig:bdpi}.  In Eq.~(\ref{fact2}) $H$ encodes calculable $m_b$ scale
contributions, $J$ encodes $\sqrt{m_b\Lambda_{\rm QCD}}$ scale contributions,
and $S$ is the non-perturbative shape function~\cite{shape}. This factorization
formula is required to describe the CLEO data on the photon energy
spectrum~\cite{CLEObsg}.

To summarize, the soft-collinear effective theory allows factorization proofs to
be simplified and formulated at the level of an effective Lagrangian and
operators. The proofs involve essentially the same steps for many exclusive and
inclusive processes. Finally, SCET provides us with a new framework for
investigating power corrections, and in the future should be used to classify
subleading non-perturbative matrix elements in a similar way to HQET.

\bigskip
This work was supported in part by the Department of Energy under the grant
DE-FG03-00-ER-41132.  I would like to thank Christian Bauer and Dan Pirjol for
enjoyable collaboration on the results presented here.

\end{document}